\documentclass[11pt]{article}

\usepackage{latexsym}
\usepackage{amssymb}
\usepackage{amsthm}
\usepackage{amscd}
\usepackage{amsmath} 
\usepackage[all]{xy}
\usepackage{graphicx}

\numberwithin{equation}{section}

\xyoption{dvips}

\CompileMatrices

\makeatletter
\renewcommand{\@makefnmark}{\mbox{\textsuperscript{}}}
\makeatother

\title{\bf The quantum black hole}
\author{Budh Ram \\
Department of Physics \\
New Mexico State University\\ 
Las Cruces, New Mexico 88003, USA\\ 
 \\
 and \\
 \\
 Prabhu-Umrao Institute of Fundamental Research \\
 A2/214 Janak Puri, New Delhi, 110058, India \\
 \\
Arun Ram \\
Department of Mathematics \\
University of Wisconsin-Madison \\
Madison, Wisconsin 53706, USA \\ \\
Nilam Ram \\
University of Virginia \\
Charlottesville, Virginia 22904, USA
}

\begin{document}

\maketitle

\begin{abstract} The quantum nature of a black hole is revealed using the 
simplest terms that one learns in undergraduate and beginning graduate
courses.  The exposition demonstrates -- vividly -- the importance and
power of the quantum oscillator in contemporary research in
theoretical physics.
\end{abstract}

\noindent {\large\bf I. INTRODUCTION}
\bigskip

``Finally, it is necessary to emphasize one major result of the whole
investigation, namely that it must be taken as well established that
the life-history of a star of small mass must be essentially different
from the life-history of a star of large mass.  For a star of small
mass the natural white-dwarf stage is an initial step towards complete
extinction.  A star of large mass ($>m$) cannot pass into the
white-dwarf stage, and one is left speculating on other
possibilities.'', wrote Chandrasekhar in the \emph{observatory}
[1,2,3] in 1934.  Thus originated the concept of a `black hole', the
collapsed state of stellar matter [4,5].

What does Einstein's theory of general relativity (GR) say about this
state of matter?

The answer was provided in a classic paper [6] by Oppenheimer and
Snyder in 1939.  They studied the collapse of a sufficiently massive
star under the idealized conditions of spherical symmetry and
pressureless stellar matter, and concluded that the star contracts
under its own gravitational attraction, its boundary $r_b$ necessarily
approaching its gravitational radius $r = \frac{2MG}{c^2}$ [now called
the Schwarzschild radius, $M$ being the mass of the star, $G$ the
Newton's constant, and $c$ the speed of light].  All energy emitted
outward from the surface of the star is reduced in escaping, by the
Doppler effect from the receding source, by the large gravitational
red-shift, $(1-\frac{r}{r_b})^{1/2}$, and by the gravitational
deflection of light which prevents the escape of radiation except
through a cone about the outward normal of progressively shrinking
aperture as the star contracts.  The star thus tends to close itself
off from any communication with a distant observer; only its
gravitation persists.  From the point of view of a distant observer,
it takes an infinite time for this asymptotic isolation to be
established; for an observer comoving with the stellar matter this
time $\tau_0$ [the proper time] is finite.  After this time $\tau_0$
an observer comoving with the matter is not able to send a light
signal from the star, as the cone within which a signal can escape
closes entirely [7].  And one says that a black hole is formed with
its surface at $r=\frac{2MG}{c^2}$, a distant observer being unable to
see inside this surface.

So what does the classical theory of general relativity tell an
observer stationed at a distance greater than $\frac{2MG}{c^2}$ about
the nature of the black hole -- the state of stellar matter inside the
sphere of radius $r=\frac{2MG}{c^2}$?  The answer is simple and
emphatic,namely, absolutely \emph{nothing}.  In other words, the
classical general relativity does not describe this collapsed state of
stellar matter -- the black hole.  Is this state describable quantum
mechanically?  The answer is in the affirmative and the purpose of the
present exposition is to give this description in the simplest terms
that one learns in undergraduate and beginning graduate physics
courses.  No new concepts are needed.  We briefly review, in the next
few sections, those concepts that are necessary and sufficient for the
tale to be told effectively.  It is guaranteed that well before the
story ends the readers would begin to see the ``silver train'' [8].

In the next section we go over (anew) the recipe that one learns to
make quantum mechanics from Newtonian mechanics.  The recipe is then
used on Eq. (34), which ensues directly from classical general
relativity, to obtain \emph{the} quantum equation, Eq. (36), which
quantizes \emph{mass} as in Eq. (50), the fundamental quantum of mass
being equal to twice the Planck mass.  The use of Bose statistics (in
section IX) \emph{reveals} the quantum nature of the black hole.

\bigskip

\noindent {\large\bf II. MAKING QUANTUM MECHANICS FROM NEWTON'S SECOND
LAW}

\medskip\noindent
Example (i)
\medskip

According to Newton's second law, the equation that describes a linear
harmonic oscillator is
$$
m\frac{d^2x}{dt^2}+kx=0,
\eqno (1)
$$ where $m$ is the inertial mass, $k$ the force constant, $x$ one of
the three Cartesian coordinates in Euclidean space, and $t$ the
Newtonian time which is the \emph{same} in all coordinate frames.
With $\omega^2 = k/m$, Eq. (1) is rewritten as
$$
\frac{d^2x}{dt^2}+\omega^2x=0.
\eqno (2)
$$ 
It has sine and cosine as solutions, which are not straight lines.
One gets straight line as a solution only when $k=0$, that is when
there is no force.

Now, to obtain a quantum mechanical equation from (1), one integrates
it to obtain
$$
\frac{1}{2}m\left(\frac{dx}{dt}\right)^2+\frac{1}{2}kx^2 = E,
\eqno (3)
$$ 
where $E$ is the constant of integration and is called the total
energy.  One then writes (3) as
$$
\frac{p_x^2}{2m}+\frac{1}{2}kx^2 = E
\eqno (4)
$$ 
with $p_x \equiv m\frac{dx}{dt}$, and replaces [9] $p_x$ by
$-i\hbar \frac{\partial}{\partial x}$ and $E$ by
$i\hbar\frac{\partial}{\partial t}$ to obtain the quantum
Schr\"odinger equation
$$
-\frac{\hbar^2}{2m}\frac{\partial^2}{\partial x^2}\psi+\frac{1}{2}
kx^2\psi 
=i\hbar\frac{\partial\psi}{\partial t}.
\eqno (5)
$$ Since the replacement $p_x\to -i\hbar\frac{\partial}{\partial x}$
is independent of $m$, one might as well put $m=1$ to obtain the
quantum equation
$$
\left(-\frac{\hbar^2}{2}\frac{\partial^2}{\partial x^2}+\frac{1}{2}
\omega^2x^2\right)
\psi
= i\hbar\frac{\partial \psi}{\partial t},
\eqno (6)
$$ 
which corresponds to the Newtonian Eq. (2).  With the
time-dependence $\psi(x,t) = \phi(x)e^{-i(E/\hbar)t}$, Eq. (6) becomes
$$
-\frac{\hbar^2}{2}\frac{d^2\phi(x)}{dx^2}
+\frac{1}{2}\omega^2x^2\phi(x) = E\phi
\eqno (7)
$$
which, when solved, gives for the energy eigenvalues [10]
$$
E_n = (n+\frac{1}{2})\hbar\omega, \ n = 0,1,2,\cdots .
\eqno (8)
$$ 
Equation (8) says, ignoring the irrelevant constant
$\frac{1}{2}\hbar\omega$, that there is a fundamental quantum of
energy equal to $\hbar \omega$, and that the state $\psi_1$ contains
just one such quantum, state $\psi_2$ two quanta, state $\psi_3$ three
quanta, and so on.  In other words, the quantum state of a harmonic
oscillator is characterised by the number of energy quanta it
contains.  The importance of Eqs. (2) and (8) will manifest itself in
section VI.

\bigskip\noindent
\emph{Example} (ii)
\smallskip

Newton's second law gives the equation ($\kappa=$ constant)
$$
m\frac{d^2\vec r}{dt^2} + \kappa\frac{\vec r}{r^3}=0
\eqno (9)
$$ 
for the so called Kepler-Coulomb problem.  In Eq. (9) the Euclidean
vector $\vec r = x\hat{i}+y\hat{j}+z\hat{k}$, and $t$ is the Newtonian
time.  To obtain the quantum equation that corresponds to (9), one
integrates (9) to get
$$
\frac{1}{2}m\left(\frac{d\vec r}{dt}\right)^2-\frac{\kappa}{r} = E
\eqno (10)
$$
(with $E$ the constant of integration) which is rewritten as
$$
\frac{\vec p^2}{2m}-\frac{\kappa}{r}=E,
\eqno (11)
$$ 
which then, using the Schr\"odinger recipe [9]: $\vec p \to
-i\hbar\vec\nabla$, $E\to i\hbar\frac{\partial}{\partial t}$,
transforms into the quantum equation
$$
\left( - \frac{\hbar^2}{2m} \vec \nabla^2 - \frac{\kappa}{r} \right)
\psi(\vec r,t) = i\hbar \frac{\partial}{\partial t} \psi(\vec r,t).
\eqno (12)
$$ 
With the time dependence $\psi(\vec r, t) = \phi(\vec r)
e^{-i(E/\hbar)t}$, Eq. (12) can be solved to give the well-known
energy eigenvalues [11]
$$E_n = -\frac{me^4}{2(4\pi\varepsilon_0)^2 \hbar^2 n^2},
\qquad n= 1,2,\ldots
$$ 
for the hydrogen atom ($\kappa = e^2/4\pi\varepsilon_0$), but for
our purpose (as will become evident later) we need the classical
kinetic energy $\vec p^2/2m$ in polar coordinates:
$$
\frac{\vec p^2}{2m} = \frac{1}{2} m \dot r^2 + \frac{L^2}{2mr^2}.
\eqno (13)
$$ 
In Eq. (13) $\dot r = \frac{dr}{dt}$ is the radial velocity and
$\vec L$ the angular momentum.  For the special case when $\vec L=0$,
Eq. (11) becomes
$$
\frac{p_r^2}{2m}-\frac{\kappa}{r} = E
\eqno (14)
$$ 
by formally defining $p_r \equiv m\dot r$.  Proper calculation of
$p^2_r$ using the Schr\"odinger prescription [12] gives for the
differential operator for $p^2_r$ the expression
$-\frac{1}{r^2}\frac{\partial}{\partial
r}\big(r^2\frac{\partial}{\partial r}\big)$, and the eigenvalue
equation that results from (14) is given by
$$
\left[
-\frac{1}{2mr^2}\frac{\partial}{\partial r}\big( r^2\frac{\partial}
{\partial r}\big)
-\frac{\kappa}{r}
\right] \psi(r) = E \psi(r).
\eqno (15)
$$ 
Since the replacement $p_r^2\to
-\frac{1}{r^2}\frac{\partial}{\partial r}\big(
r^2\frac{\partial}{\partial r}\big)$ is independent of $m$ in $p_r =
m\dot r$, we again put $m=1$ in (15) to obtain the quantum equation
$$
\left[
-\frac{1}{2r^2}\frac{\partial}{\partial r}\big( r^2\frac{\partial}
{\partial r}\big)
-\frac{\kappa}{r}
\right] \psi(r) = E \psi(r)
\eqno (16)
$$ 
which corresponds to the Newton equation (9) with $m=1$ in it.
Eq. (16) will show its importance in section V.
\bigskip

\noindent {\large\bf III. STRAIGHT LINES IN MINKOWSKI SPACE}
\medskip

With the proclamation: ``Henceforth space by itself and time by itself
are doomed to fade away into mere shadows, and only a kind of union of
the two will preserve an independent reality.'' by Minkowski [13],
originated his 4-dimensional spacetime, now simply called
\emph{Minkowski space} in which the distance $ds$ between two adjacent
world points (events) is given by [14]
$$
ds^2 = dt^2 - dx^2-dy^2-dz^2.
\eqno (17)
$$ 
The \emph{proper} time interval $d\tau$ between two neighboring
points for which $dx=dy=dz=0$ is then given by
$$
d\tau^2=ds^2=dt^2-dx^2-dy^2-dz^2,
\eqno (18)
$$
or
$$
d\tau = (1-u^2)^{1/2}dt,
\eqno (19)
$$ 
and is an \emph{invariant} quantity.  The relation (17) is
essentially non-Euclidean [15].  Nevertheless four-vectors like
$x^\mu$ ($\mu=0,1,2,3$), $\frac{dx^\mu}{d\tau}$ (4-velocity),
$\frac{d^2x^\mu}{d\tau^2}$ (4-acceleration), $p^\mu =
m\frac{dx^\mu}{d\tau}$ (4-momentum), $\frac{dp^\mu}{d\tau}$,
$\kappa^\mu$ (Minkowski 4-force) are constructed [16].  And special
relativistic extension of Newton's second law takes the form
$$
\frac{dp^\mu}{d\tau} = \kappa^\mu,
\eqno (20)
$$
or
$$
m\frac{d^2x^\mu}{d\tau^2} = \kappa^\mu.
\eqno (21)
$$
When $\kappa^\mu=0$, only then
$$
\frac{d^2x^\mu}{d\tau^2}=0
\eqno (22)
$$
\emph{irrespective of what the value of $m$ is}.

The solutions of Eq. (22) are straight lines.  Equation (22) then
becomes the statement of Newton's first law.  It is to be noted that
Eq. (21) can never have straight lines as their solutions, \emph{even
in Minkowski space}, as long as $\kappa^\mu\ne 0$.
\bigskip

\noindent {\large\bf IV. CASTING NEWTON'S SECOND LAW IN THE `FORM' OF
HIS FIRST LAW} 
\medskip

But in the case when the force is Newton's gravitational force,
something very remarkable happens.  In this case Newton's second law
says:
$$
m\frac{d^2\vec r}{dt^2}+\frac{GmM\vec r}{r^3} = 0.
\eqno (23)
$$ 
The remarkable thing is that the $m$ in the first term, usually
called the inertial mass, is the same as the $m$ in the second term,
usually called the gravitational mass.  And Eq. (23) reduces to
$$
\frac{d^2\vec r}{dt^2} + \frac{GM\vec r}{r^3}=0,
\eqno (24)
$$
which in Minkowski space takes the form
$$
\frac{d^2x^i}{d\tau^2}+\frac{GMx^i}{r^3} = 0
\qquad(i=1,2,3).
\eqno (25)
$$ 
The remarkable feature of Eq. (25) is that it is \emph{independent}
of $m$ just as Eq. (22).  Note that Eq. (25) does not have straight
lines as its solutions, whereas Eq. (22) does.  Now the question
arises whether one can cast Eq. (25) in a new form that has as its
solutions, or curves which a world point traces, straight lines.  The
answer is a definite yes.  All one has to do is go to what is called
4-dimensional Riemannian space [17] in which the distance $ds$, or the
line element (as it is sometimes called), is given by
$$
ds^2 = g_{\mu\nu}dx^\mu dx^\nu\qquad (\mu,\nu=0,1,2,3)
\eqno (26)
$$ 
with $g_{\mu\nu}(x^\mu)$ functions of $x^\mu$, and the equations of
geodesics, the \emph{straightest lines} [18], are
$$
\frac{d^2x^\mu}{d\tau^2} + \Gamma_{\nu\lambda}^\mu\frac{dx^\nu}{d\tau}
\frac{dx^\lambda}{d\tau} = 0.
\eqno (27)
$$ 
In Eqs. (27), $\tau$ is the \emph{proper time}, and the affine
connections $\Gamma_{\nu\lambda}^\mu$ are given by
$$
\Gamma_{\nu\lambda}^\mu = \frac{1}{2}g^{\sigma\mu}
\left\{ \frac{\partial g_{\lambda\sigma}}{\partial x^\nu}
+\frac{\partial g_{\nu\sigma}}{\partial x^\lambda}
-\frac{\partial g_{\lambda\nu}}{\partial x^\sigma}
\right\}.
\eqno (28)
$$
Thus once the $g_{\mu\nu}$ are determined from Einstein equation [17]
$$
R^{\mu\nu}-\frac{1}{2}g^{\mu\nu}R = -8\pi T^{\mu\nu}
\eqno (29)
$$ 
for a given distribution of matter, the $\Gamma^\mu_{\nu\lambda}$
are obtained using (28), and the geodesic equations (27) are solved to
obtain the geodesics or the ``straightest world lines''.

The importance of Eqs. (27) lies in the fact that it was these
equations which Einstein solved, though using approximation methods,
to obtain the correct value for the advance of perihelion of mercury
[19].  And it is precisely the integral of these equations which we
are going to use in the next section for the special spherically
symmetric case.  The first integral of Eqs. (27) for time-like
geodesics is
$$
g_{\mu\nu}\frac{dx^\mu}{d\tau}\frac{dx^\nu}{d\tau} =1.
\eqno (30)
$$
\bigskip

\noindent {\large\bf V. MAKING QUANTUM MECHANICS FROM TIME-LIKE GEODESICS}
\medskip

The exact, static, spherically symmetric solution of Einstein
equations for {\it vacuum},
$$
R^{\mu\nu}=0,
\eqno (31)
$$
was obtained by Karl Schwarzschild [20] in 1916 and is given by 
$$
ds^2 = (1-(2M/r))dt^2 -
\left(\frac{dr^2}{1-(2M/r)}+r^2(d\theta^2+\sin^2\theta d\phi^2)\right). 
\eqno (32)
$$ 
Note that Eq. (32) ensures that for $r\to \infty$, it goes to the
Minkowski line element
$$
ds^2 = dt^2 - (dr^2+r^2(d\theta^2+\sin^2\theta d\phi^2)).
\eqno (33)
$$ 
In (32) $M$ is the Newtonian point mass located at the origin of
coordinates $r, \theta, \phi$.  We want to emphasize that in solving
Eqs. (31) $M$ arises as a constant of integration and is identified
with the Newtonian mass only when one falls back to the Newton's
gravitational law according to which the gravitational potential at a
distance $r$ from a point mass $M$ is $-GM/r$.  In other words, mass
or gravitation arises or is \emph{created} out of the vacuum [21].
Put succinctly, Einstein equations (31) manifest only gravitation
(mass) and \emph{nothing else}.

However, with the expressions for $f_{tt}$, $f_{rr}$,
$f_{\theta\theta}$ and $f_{\phi\phi}$ from (32), when the time-like
geodesic Eqs. (27) are integrated there arise two more constants of
integration [22] designated by $E$ and $L$, which, with the postulate
[23] that test particles follow the time-like geodesic, are
respectively identified with the energy per unit mass and angular
momentum per unit mass of the test particle.  But since we are
interested only in pure gravitation ($M$) and \emph{nothing else}, the
values of these constants in our case of interest are
\emph{necessarily zero}.  As a result only \emph{one} equation remains
     [24], namely
$$
\frac{1}{2}\left(\frac{dr}{d\tau}\right)^2-\frac{M}{r} = -\frac{1}{2}
\eqno (34)
$$ 
which is characterized by only one parameter, the Newtonian mass
$M$ located at $r=0$, and describes the state of pure gravitation in
the region [25] $r\le 2M$, the first term being zero at $r=2M$, the
Schwarzschild radius.  Bigger the value of $M$, larger is the
Schwarzschild radius.

It should be apparent now to the reader, in view of section II and the
fact that the proper time $\tau$ has the same character as that of
Newton's time $t$, how to make the quantum equation from (34): simply
by replacing $p_r^2 = \dot r^2 = (dr/d\tau)^2$ by the differential
operator $\displaystyle{-\frac{1}{r^2}\frac{\partial}{\partial
r}\left(r^2\frac{\partial}{\partial r}\right) }$.  And one gets [26]
$$
-\frac{1}{2r^2}\frac{\partial}{\partial r}\left(r^2\frac{\partial}
{\partial r}\right)\psi
-\frac{M}{r}\psi = -\frac{1}{2}\psi
\eqno (35)
$$ 
as the quantum equation which corresponds to the classical equation
(34).  It is obvious that Eq. (35) describes a bound state of binding
energy $1/2$.  With $U=r\psi$ and $M=\mu/4$, Eq. (35) reduces to the
simple form
$$\left(-\frac{1}{2}\frac{d^2}{dr^2} - \frac{\mu/4}{r}\right)U
= -\frac{1}{2}U.
\eqno (36)
$$
We shall return to Eq. (36) in section VII.
\bigskip

\noindent {\large\bf VI. EMERGENCE OF PHOTONS FROM MAXWELL'S EQUATIONS}
\bigskip

Maxwell's equations for pure radiation (light waves) can be expressed
in terms of only the spatial part $\vec A$ of the Minkowskian
4-potential $A_\mu$ ($A_0 = \phi = 0$), and takes the form
$$
\nabla^2\vec A - \frac{d^2 \vec A}{dt^2}=0, \eqno (37) 
$$
$$ 
\vec\nabla\cdot \vec A = 0, \hspace*{1cm} 
\eqno (38)
$$ $\vec A$ being defined at all world points [27].  With
\emph{periodic} boundary conditions on $\vec A$, the general solution
of (37) can be represented as a series of orthogonal `eigenvalues':
$$
\vec A = \sum_\lambda q_\lambda(t)\vec A_\lambda(\vec r)
\eqno (39)
$$ 
where $\vec A_\lambda$ depends only upon the space coordinate and
$q_\lambda$ only on the time coordinate.  $\vec A_\lambda$ of course
obeys the periodic boundary conditions.  The $\vec A_\lambda$ then
satisfy the wave equation
$$
\nabla^2\vec A_\lambda + K_\lambda^2\vec A_\lambda = 0,
\eqno (40)
$$
with
$$
\vec \nabla\cdot \vec A_\lambda = 0,
\eqno (41)
$$
and $q_\lambda$ satisfy the Newton's equation for a harmonic oscillator
$$
\frac{d^2q_\lambda}{dt^2} + \omega_\lambda^2q_\lambda = 0,
\qquad\hbox{for each wavelength $\lambda$.}
\eqno (42)
$$ 
Thus arises the harmonic oscillator naturally from Maxwell's
equations; and one says that pure radiation is composed of them
[harmonic oscillators].  Eq. (42), when quantized in the manner of
section II, gives for energy eigenvalues
$$
E_\lambda = (n_\lambda+\frac{1}{2})\hbar\omega_\lambda,
\qquad n_\lambda = 0,1,2,\ldots,
\eqno (43)
$$ 
and one says that a state $\vert \ n_{\lambda_1}, n_{\lambda_2},
\ldots \rangle$ of pure radiation is characterized by the numbers
$n_{\lambda_1}, n_{\lambda_2},\ldots$ of light quanta (photons) of
respective energies $\hbar\omega_{\lambda_1},
\hbar\omega_{\lambda_2},\ldots $.
\bigskip

\noindent {\large\bf VII. EMERGENCE OF PAIRED QUANTA OF PURE
GRAVITATION}
\bigskip

The radial Schr\"odinger equation for the $N'$-dimensional Coulomb or
Kepler problem can be written [28,29] as
$$
\left[ -\frac{1}{2}\big( \frac{d^2}{dr^2}-
\frac{(\ell'+N'/2-3/2)(\ell'+N'/2-1/2)}{r^2}\big)-\frac{\alpha}{r}
\right] U(r) = -BU(r),
\eqno (44)
$$ 
where the orbital quantum number $\ell'$ is a positive integer or
zero and $B$ the Coulomb or Kepler binding energy.  In $N$ dimensions
the oscillator counterpart of Eq. (44) is
$$
\left[ -\frac{1}{2}\big( \frac{d^2}{ds^2}-
\frac{(\ell+N/2-3/2)(\ell+N/2-1/2)}{s^2}\big)+\frac{1}{2}\omega^2s^2
\right]\phi_{n\ell}^N(s)
=(2n+\ell+N/2)\omega\phi_{n\ell}^N(s),
\eqno (45)
$$
where $n=0,1,\ldots,$ and $\ell$ the corresponding angular momentum
quantum number. Under the transformation [29,30,31]
$$
r = \rho^2, \qquad\hbox{and}\qquad U = \rho^{1/2}\phi,
\eqno (46)
$$
Eq. (44) becomes
$$
\left[ -\frac{1}{2}\left( \frac{d^2}{d\rho^2}- 
\frac{(2\ell'+(2N'-2)/2-3/2)(2\ell'+(2N'-2)/2-1/2)}{\rho^2}\right)+
4B\rho^2\right]
\phi(\rho) = 4\alpha \phi(\rho).
\eqno (47)
$$ 
Comparison of Eqs. (45) and (47) shows that there is a satisfactory
mapping when [31]
$$
\left.
\begin{array}{cc}
\ell = 2\ell', &N=2N'-2, \\
\frac{1}{2}\omega^2 = 4B, &(2n+\ell+N/2)\omega = 4\alpha,
\end{array}\right\}.
\eqno (48)
$$

Now we go back to Eq. (36) and note that it [Eq. (36)] is Eq. (44) with
$$
\ell'=0,\ \ N'=3,\ \ B=1/2,\ \ \alpha = \mu/4;
\eqno (49)
$$ 
and hence it represents a 4-dimensional harmonic oscillator with
angular momentum $\ell=0$, angular frequency $\omega=2$, and $\mu$
given by
$$
\mu_n = 2(n+1)\omega,\qquad n= 0,1,2,\ldots.
\eqno (50)
$$
Thus the oscillator {\it shows} itself again.  Though this time it is
a four-dimensional one, as 
opposed to the one dimensional one in the case of pure radiation (see
section VI) [32].

Equation (50) says that the fundamental quantum of \emph{pure}
gravitation is of frequency $2$ or of mass (energy) twice the Planck
mass (energy) 
[33].  But the quanta always come in pairs.  And one says that the
state $\vert n\rangle$ of pure gravitation is characterized by $n$
pairs of mass quanta.
\bigskip

\noindent {\large\bf VIII. BLACK HOLE ENTROPY}
\bigskip

After the classic paper of Oppenheimer and Snyder [6] in 1939, the
field of general relativity and black holes was dormant for about
twenty years.  Then there was a burst of activity.  In 1963 Roy Kerr
[34] found another exact, stationary, axisymmetric, solution of
Einstein equations, characterised by two parameters that are
identified with the mass and angular momentum of the socalled Kerr
inner region [25].  During the 1970s, the mathematics of black holes
was cast in a form that closely resembled that of thermodynamics [35].
In other words four laws of black hole mechanics were formulated
which were analogous to the four laws of thermodynamics, the
quantities $M$ (mass), $\kappa$ (surface gravity), $A$ (area) of a
black hole being analogous, respectively, to the thermodynamic
quantities $E$ (energy), $T$ (temperature), $S$ (entropy).  However,
Bekenstein [36], based on his calculations and gedanken experiments,
went beyond a mere analogy.  He proposed that the area $A$ of a black
hole represents in actuality the \emph{physical} entropy $S$ of a
black hole, it being proportional to $A$ ($S=\alpha A$).  But nobody
believed him [37].  Then came Hawking [38] with his semi-classical
calculation of scattering of a quantized scalar field of the boundary
of a classical Schwarzschild black hole, and \emph{his} interpretation
of the results: \emph{that a black hole emits particles}, the emission
spectrum being exactly the same as that of a black body of temperature
$T = \kappa/2\pi$. [$\kappa = 1/4M$ for a Schwarzschild black hole
giving its temperature $T=1/(8\pi M)$.]  This fixed the
proportionality constant $\alpha$ to be $1/4$ giving the physical
entropy $S= A/4$ for a black hole.

To obtain $S = A/4$ (the socalled Bekenstein-Hawking relation)
theoretically has been a \emph{challenge} for theoretical and
mathematical physicists ever since 1975.  Many a calculation have been
done using a variety of mathematical techniques [39].  But \emph{none}
of these calculations has the \emph{simplicity} and \emph{cleanliness}
possessed by the one given in the next section, a calculation that
\emph{can} be used in the classroom. 
\bigskip

\noindent {\large\bf IX. CALCULATION OF BLACK HOLE ENTROPY A LA BOSE}
\bigskip

Historically it was Planck who, in view of Kirchhoff's theorem, used
the one-dimensional oscillators (as the simplest mathematical
convenience) with the discrete energy element $\varepsilon = h\nu$,
$\nu$ being the frequency of the oscillators, to derive his black body
radiation law [40], Einstein later interpreting $\varepsilon = h\nu$
as the energy quantum of light itself [40,41].  However, the
connection of a light quantum with the energy element of an oscillator
was not formally established till after the advent of quantum
mechanics in 1925 and its application to pure radiation, as described
briefly in section VI.  Despite this, Bose [42] in 1924 used the light
quanta, coupled with the fact that they are indistinguishable from one
another, to derive the entropy [along with the Planck spectral
law] of pure radiation within a black body.  Now that the connection
of the fundamental energy element of a four-dimensional oscillator
with the mass quantum of pure gravitation has been rigorously shown
quantum mechanically, as described in section VII, it only behooves that
we use Bose's method to calculate the entropy $S$ of pure gravitation 
within a black hole.  We
could simply use the formula for the entropy given in Bose's paper,
but we rederive it below so that it is expressed in terms of the area
$A$ of the black hole.

We saw in section VII that the quantum of pure mass is of $\omega=2$
and that these mass (energy) quanta always come in pairs.  So let the
mass of a Schwarzschild black hole as in the paper of Oppenheimer and
Snyder [6] be $M_s$ and let it contain $2N$ quanta.  Then $M_s =
2N\omega$.  Let us rewrite Eq. (50) as
$$
\epsilon_n = \frac{\mu_n}{2} = (n+1)\omega,
\qquad n= 0,1,2,\ldots,
\eqno (51)
$$ 
so that the $n$th $\epsilon$-state contains $n$ quanta.  The 
thermodynamic quantity $E$ in Bose's paper [42] is given by
$$
E = \frac{M_s}{2} = N\omega.
\eqno (52)
$$ 
This then allows us to literally apply Bose's method to calculate
the thermodynamic probability of the macroscopically defined state
$(N,E)$.

Let $p_0$ be the number of vacant $\epsilon$-cells, $p_1$ the number
of those cells which contain one quantum, $p_2$ the number of
$\epsilon$-cells containing two quanta, and so on.  Then the
probability of the state defined by the $p_r$ is
$$
W = \frac{P!}{p_0!p_1!\cdots},
\eqno (53)
$$
where
$$
P = \sum_{r} p_r
\eqno (54)
$$
is the total number of $\epsilon$-cells over which
$$
N = \sum_r rp_r
\eqno (55)
$$ 
quanta are distributed.  Since $p_r$ are large numbers we have,
using Stirling's formula,
$$
\ln W = P\ln P - \sum_r p_r\ln p_r.
\eqno (56)
$$ 
Then it is straightforward [42], by using the method of Lagrange
multipliers, to maximize (56) satisfying the auxiliary conditions
(52) and (55), and obtain
$$
p_r = P(1-e^{-\omega/\beta})e^{-r\omega/\beta}, 
\eqno (57)
$$
$$
N = P(e^{\omega/\beta}-1)^{-1}, \hspace*{1cm}
\eqno (58)
$$
and
$$
S = \frac{E}{\beta} - P\ln(1-e^{-\omega/\beta}).
\eqno (59)
$$ 
From the condition $\frac{\partial S}{\partial E} = \frac{1}{T}$,
one obtains $\beta = T$; and (59) becomes
$$
S = \frac{E}{T} - P\ln(1-e^{-\omega/T}).
\eqno (60)
$$
As mentioned in section VIII, for a Schwarzschild black hole $T=
\frac{1}{8\pi M_s}$ and $A=16\pi M_s^2$, and (60) becomes [43] 
$$
S = \frac{A}{4} - \frac{M_s}{4}(e^{A/M_s}-1)\ln(1-e^{-A/M_s}),
\eqno (61)
$$
which, for large $M_s$, physically tends to 
$$
S = \frac{A}{4},
\eqno (62)
$$
\emph{same} as the Bekenstein-Hawking relation.
\bigskip

\noindent {\large\bf X. THE REVELATION}
\bigskip

So what \emph{is} the quantum nature of a black hole?

A black hole is a Bose-Einstein ensemble of quanta of mass equal to
twice the Planck mass, confined in a sphere of radius twice the black
hole mass.

\vspace{1.5cm}

The authors are grateful to R.S. Bhalerao for a critical reading of
the manuscript.

\newpage

\noindent {\large\bf REFERENCES AND FOOTNOTES}
\bigskip
\begin{enumerate}
\item[{[1]}] S. Chandrasekhar, ``Stellar Configurations with
degenerate cores'', The Observatory \textbf{57} 373--377 (1934). 

\item[{[2]}] For an historical recollection of the steps leading to
the above statements, see S. Chandrasekhar, ``The Richtmyer Memorial
Lecture -- Some Historical Notes'', Am.\ J.\ Phys.\ \textbf{37}
577--584 (1969).

\item[{[3]}] The limiting mass $m$ is the so called Chandrasekhar limit.

\item[{[4]}] It was L.\ Landau, [ ``On the theory of stars'', Phys.\
Z.\ Soviet \textbf{1}, 285--287 (1932); reprinted in \emph{Neutron
Stars, Black Holes and Binary X-ray sources}, eds. H.\ Gursey and R.\
Ruffini, D.\ Reidel Publishing Co., Boston, 1975, pp.\ 271--273.] who
stated that a star with mass $>m$ would collapse but did not think
that such a state could exist in reality.

\item[{[5]}] The word black hole to describe such a collapsed state
was coined in 1968 by J.A.\ Wheeler [``Our Universe: the known and the
unknown'', American Scientist \textbf{56}, 1--20, 1968].

\item[{[6]}] J.R.\ Oppenheimer and H.\ Snyder, ``On continued
gravitational contraction'', Phys.\ Rev.\ \textbf{56}, 455--459
(1939).

\item[{[7]}] These conclusions are speculation free as they are
founded on solid mathematical formulas obtained using Einstein's
equations of GR.

\item[{[8]}] J.S. Rigden, ``Editorial: The fallacy of immediacy'',
Am. J. Phys. \textbf{55}, 395 (1987); B.\ Ram,
``The silver train'', Am.\ J.\ Phys. \textbf{55}, 968 (1987).

\item[{[9]}] J.J.\ Prentis and W.A.\ Fedak, ``Energy Conservation in
Quantum Mechanics'', Am.\ J.\ Phys.\ \textbf{72}, 580--590 (2004).

\item[{[10]}] See, for example, L.I.\ Schiff, ``Quantum mechanics'',
3rd edition (McGraw-Hill, New York, 1968), Chap. 4.

\item[{[11]}] See, for example, R.\ Eisberg and R.\ Resnick,
\emph{Quantum Physics of Atoms, Molecules, Solids, Nuclei, and
Particles} (Wiley, New York, 1985), chap.\ 7.

\item[{[12]}] In this connection we draw the reader's attention to page
125 of W.\ Yourgrau and S.\ Mandelstam, \emph{Variational Principles
in Dynamics and Quantum Theory}, 3rd ed.\ (W.A.\ Saunders Co.,
Philadelphia, 1968).

\item[{[13]}] H.\ Minkowksi, ``Space and time'', in \emph{The
Principle of Relativity} (Dover Publications, New York, 1923) p. 75.

\item[{[14]}] From now on we use units in which $\hbar$ (Planck's
constant) $=c$ (velocity of light) = G (Newton's gravitational
constant) $=k$ (Boltzmann constant) $=1$.

\item[{[15]}] See, for example, D.F.\ Lawden, \emph{An Introduction to
Tensor calculus, Relativity and Cosmology}, 3rd ed.\ (Dover
Publications, New York, 2002), p.\ 15.

\item[{[16]}] Ref.\ 15, chap.\ 3.

\item[{[17]}] Ref.\ 15, chaps.\ 5 and 6.

\item[{[18]}] A geodesic is a curve which has the property that the
tangents at all its points are parallel.  In Euclidean space such a
curve is, of course, called a straight line.

\item[{[19]}] In this connection, see L.I.\ Schiff, ``On experimental
tests of the general theory of relativity'', Am.\ J.\ Phys.\
\textbf{28}, 340--343 (1960).

\item[{[20]}] K.\ Schwarzschild, ``\"Uber das Gravitationsfeld eines
Massenpunkte nach der Eisensteinschen Theorie'',
Sitzber. Deut. Akad. Wiss. Berlin, KI.\ Math.--Phys.\ Tech., 189--196
(1916); for an English translation see arXiv:physics/9905030.

\item[{[21]}] For steps leading to this, see Ref.\ 15, pages 142-146.

\item[{[22]}] See Ref. 15, pp. 147--148.

\item[{[23]}] A.\ Einstein, \emph{The meaning of relativity}, 5th ed.\
(Princeton University Press, 1974), p.\ 79.

\item[{[24]}] With the Schwarzschild solution (32), Eq. (30), the
first integral of geodesic equations (27), takes the form
$\frac{1}{2}\dot r^2 + \frac{L^2}{2r^2}-\frac{M}{r}-\frac{ML^2}{r^3}
=\frac{1}{2}(E^2-1)$, which, with $E=L=0$, reduces to Eq. (34).

\item[{[25]}] Following S.\ Chandrasekhar, \emph{Truth and Beauty} 
(University of Chicago Press, 1987), p. 71, we shall call this the
\emph{inner} region.

\item[{[26]}] Eq. (35) was first obtained in B.\ Ram, ``The mass
quantum and black hole entropy'', Phys.\ Lett.\ A \textbf{265}, 1--4
(2000).

\item[{[27]}] For details see, W.\ Heitler, \emph{The Quantum Theory
of Radiation}, 3rd ed.  (Dover, 1984), \S\S 6 and 9.

\item[{[28]}] A.\ Chatterjee, ``Large-$N$ expansion in quantum
mechanics'', Phys.\ Rep.\ \textbf{186}, 249--370 (1990).

\item[{[29]}] D.S.\ Bateman, C.\ Boyd and B.\ Dutta-Roy, ``The mapping
of the Coulomb problem into the oscillator,'' Am.\ J.\ Phys.\
\textbf{60}, 833--836 (1992).

\item[{[30]}] H.A.\ Mavromatis, ``Mapping of the coulomb problem into
the oscillator: A comment on the papers by Bateman \emph{et al}. [Am.\
J.\ Phys.\ \textbf{60}(9), 833--836 (1992)], and P.\ Pradhan [Am.\ J.\
Phys.\ \textbf{63}(7), 664 (1995)]'', Am.\ J.\ Phys.\ \textbf{64},
1074--1075 (1996).

\item[{[31]}] H.A.\ Mavromatis, ``A straightforward mapping of the
arbitrary dimensional Coulomb problem into the isotropic oscillator'',
Rep.\ Math.\ Phys.\ \textbf{40}, 17--19 (1997).

\item[{[32]}] For the Kerr (Ref.\ 34) solution of Einstein equations,
the first integral, Eq. (30), of time-like geodesics is given by
$$
\frac{1}{2}\dot r^2 -
\frac{M}{r}+\frac{1}{2}(1-E^2)(1+a^2/r^2)+\frac{L^2}{2r^2} 
-\frac{M}{r^3}(L-aE)^2 = 0,
\eqno (K1)
$$
which, with $E=L=0$, reduces to
$$
\frac{1}{2}\dot r^2-\frac{M}{r}+\frac{a^2}{2r^2} = -\frac{1}{2}.
\eqno (K2)
$$ 
In the above two equations $a$ is angular momentum of the Kerr
inner region.  It is left as an exercise for the reader to obtain the
quantum equation that corresponds to (K2) and then obtain the
corresponding expression (analogous to Eq. (50)) for the eigenvalues
of the mass $\mu$.  (For hint, see B.\ Ram, ``The mass quantum and
black hole entropy II'', arXiv: gr-qc/0101056).

\item[{[33]}] One Planck mass $\displaystyle{ (m_p) =
\left(\frac{\hbar c}{G}\right)^{1/2} }$, one Planck energy $=m_pc^2$.

\item[{[34]}] R.P.\ Kerr, ``Gravitational field of a spinning mass as an
example of algebraically special metrics'', Phys.\ Rev.\ Lett.\
\textbf{11}, 237--238 (1963).

\item[{[35]}] J.M.\ Bardeen, B.\ Carter, and S.W.\ Hawking, ``The four
laws of black hole mechanics'', Commun.\ Math.\ Phys.\ \textbf{31},
161--170 (1973).

\item[{[36]}] J.D.\ Bekenstein, ``Black holes and entropy'', Phys.\
Rev.\ D7, 2333-2346 (1973).

\item[{[37]}] J.D.\ Bekenstein, ``Do we understand black hole
entropy?'', arXiv:gr-qc/9409015.

\item[{[38]}] S.W.\ Hawking, ``Particle creation by black holes'',
Commun.\ Math.\ Phys.\ \textbf{43}, 199-220 (1975).

\item[{[39]}] For a partial list of references on these calculations
see C.\ Rovelli, ``Strings, loops and others: a critical survey of the
present approaches to quantum gravity'', arXiv: gr-qc/9803024, and T.\
Damour, ``The entropy of black holes: a primer'', arXiv:
hep-th/0401160.

\item[{[40]}] M.J.\ Klein, ``Max Planck and the beginning of the
quantum theory'', Arch.\ Hist.\ Exact.\ Sci.\ \textbf{1}, 459--479
(1962).

\item[{[41]}] A.B.\ Arons and M.B.\ Peppard, ``Einstein's proposal of
the photon concept, a translation of the Annalen der Physik paper of
1905'', Am.\ J.\ Phys.\ \textbf{33}, 367--374 (1965).

\item[{[42]}] Bose, ``Plancks Gesetz und Lichtquanten hypothese'', Z.\
Phys.\ \textbf{26}, 178-181 (1924); English translation in O. \
Theimer and B.\ Ram, ``The beginning of quantum statistics'', Am.\ J.\
Phys.\ \textbf{44}, 1056--1057 (1976).

\item[{[43]}] Eq. (61) was first obtained in B.\ Ram, ``The mass
quantum and black hole entropy'', Phys.\ Lett.\ A \textbf{265}, 1--4
(2000).
\end{enumerate}

\end{document}